\documentclass[aps,twocolumn,epsfig]{revtex4}
\newcommand \beq{\begin{eqnarray}}
\newcommand \eeq{\end{eqnarray}}
\newcommand \ga{\raisebox{-.5ex}{$\stackrel{>}{\sim}$}}
\newcommand \la{\raisebox{-.5ex}{$\stackrel{<}{\sim}$}}
\usepackage{graphicx}

\begin{document}

\title{Stability of trapped fermionic gases with attractive interactions}
\author{Benjamin M. Fregoso and Gordon Baym}
\date{\today}
\affiliation{Department of Physics, University of Illinois at
Urbana-Champaign, 1110 West Green Street, Urbana, Illinois
61801 }

\begin{abstract}

    We present a unified overview, from the mean-field to the unitarity
regime, of the stability of a trapped Fermi gas with short range attractive
interactions.  Unlike in a system of bosons, a Fermi gas is always stable in
these regimes, no matter how large the particle number.  However, when the
interparticle spacing becomes comparable to the range of the interatomic
interactions, instability is not precluded.

\end{abstract}
\maketitle

\section{Introduction}

    The freedom to tune the interparticle interactions in dilute Fermi and
Bose gases using Feshbach resonances raises the question of the stability of
trapped atomic gases with attractive interactions against collapse.  Weakly
interacting Bose gases with attractive interactions become mechanically
unstable once the particle number exceeds a critical value, of order the ratio
of the oscillator length to the magnitude of the s-wave scattering length
\cite{hulet95,ruprecht,baym,erich}.  The question is whether Fermi systems are
always stabilized, and by what mechanisms, or can they ever be similarly
unstable?  Our aim in this paper is to present an overview of the problem in
the different parameter regimes, putting together and extending the various
arguments that have been given for the stability (and instability) of Fermi
systems \cite{gehm,henning,menotti,baker,carlson,ohara,bourdel,
partridge2,partridge,erich2,stoof,houbiers,pethick2} within a unified
framework.

    At first, one might think that the Fermi energy of the atoms would be
sufficient to support a trapped Fermi gas against collapse, in the way, for
example, that electron degeneracy pressure supports white dwarfs against
gravitational collapse.  In a white dwarf the gravitational energy scales as
the inverse of the system radius, ${\cal R}$, while the Fermi energy scales as
${\cal R}^{-2}$ non-relativistically; instability sets in only in the
relativistic limit, when the Fermi energy begins to scale as $\sim {\cal
R}^{-1}$.  In a trapped Fermi gas with short range s-wave interactions between
different components, the Fermi energy similarly scales as ${\cal R}^{-2}$,
while the interaction energy scales, within mean-field theory, as ${\cal
R}^{-3}$.  Thus the Fermi energy is not, in itself, adequate to prevent an
instability.

\begin{figure}[h]
\centering
\includegraphics[width=0.37\textwidth]{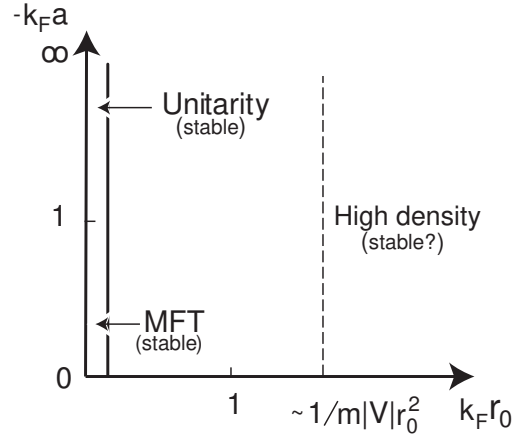}
\caption{
Regions of stability in the $k_F r_0$, $k_F |a|$ plane of a two component
Fermi gas with short range interactions.  Here $k_F$ is the Fermi momentum,
$a$ the s-wave scattering length, and $r_0$ the range of the interactomic
potential.  }
\label{fig:ka_kr}
\end{figure}

    In the Fermi problem with low energy s-wave interactions, one must
distinguish several regimes.  The first is that of the weak interactions where
the s-wave scattering length, $a$, is small compared with the interparticle
spacing, and mean-field theory is valid.  Calculating within mean-field theory
one finds an instability at high density \cite{stoof,houbiers,pethick2, gehm},
$k_F |a| > \pi/2$, or large particle number $N^{1/6} \ga 0.6 d/|a|$
\cite{henning,menotti}, where $k_F$ is the Fermi momentum, and
$d=\sqrt{\hbar/m\omega}$ is the oscillator length, with $\omega$ the
oscillator frequency (assumed isotropic); however, at such densities the
interparticle spacing is comparable to the scattering length, and the
mean-field calculation becomes invalid.  Indeed, in the mean-field regime
($k_F|a|<<1$) the system is always stable.  The second -- ``unitary" -- regime
is when the scattering length is large compared with the interparticle
spacing, itself large compared with the range of the interparticle potential.
Here the system is again always stable, since the total energy remains
positive
\cite{henning,gehm,baker,carlson,ohara,bourdel,partridge2,partridge,erich2}.
The third regime is when the interparticle spacing becomes comparable to the
range, $r_0$, of the interparticle potential.  In this regime the system can
in principle become unstable.  Figure \ref{fig:ka_kr} schematically
illustrates these regimes in the $k_F r_0$, $k_F |a|$ plane.

    We first study the stability of the Fermi gas in mean field theory,
working in the local-density approximation in which the gas is assumed at each
point to be locally homogeneous \cite{butts}, and derive analytically the
mean-field limits on the density and particle number.  As we see, the two
instability conditions, $k_F |a| > \pi/2$, and number $N^{1/6} \ga 0.6 d/|a|$,
are equivalent.  We next discuss the unitary regime.  By means of a simple
interpolation formula for the density dependence of the energy, we show that
the gas should be stable, with no limit in the number of particles in the
system.  We show, by using the Hartree-Fock approximation as a rigorous
upper bound to the total energy of a system with an attractive square well
interparticle interaction, how stability can break down in the high
density regime.

\section{Stability in mean-field}

    We consider a Fermi gas with equal numbers of two components (denoted by
$\uparrow$ and $\downarrow$) in equilibrium at zero temperature in an
isotropic harmonic trap, $V(r)= \frac{1}{2}m\omega^2 r^2$.  The energy of the
system in the local density approximation is
\begin{eqnarray}
  E &=& \frac{3\hbar^2}{10m}(3\pi^2)^{2/3}\int d{\mathbf r}~n^{5/3} \nonumber \\
   &&+ \frac{1}{2}m\omega^2 \int d{\mathbf r}~ r^2~ n  + g \int d{\mathbf r}~
  n_{\uparrow}n_{\downarrow},
  \label{e_func}
\end{eqnarray}
where $g = 4\pi\hbar^2 a/m$ is the coupling constant, and
$n=2n_{\uparrow}=2n_{\downarrow}$ is the total density of particles.  The
short range s-wave interactions between particles of the same species
essentially vanish due to the antisymmetry of the many-fermion wave function.

\begin{figure}
\includegraphics[width=0.47\textwidth]{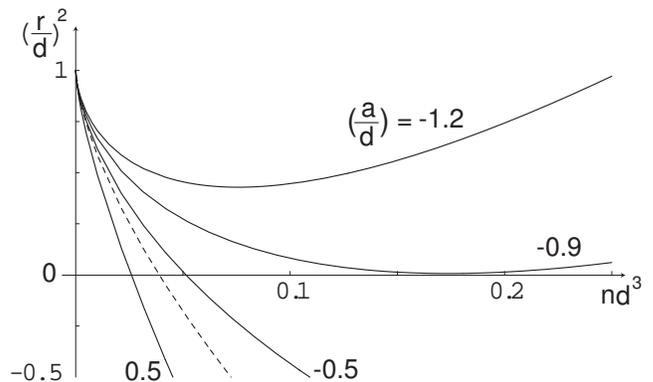}
\caption{
Distance from center of trap, squared, as a function of the density, $n$,
Eq.~(\ref{rn}), for the value $R$ = $d$. For large negative $a$ the curve
does not intersect the $n$-axis, and the system becomes unstable.  The dashed
curve shows the characteristic radius squared vs. density calculated with the
effective scattering length, Eq.~(\ref{aeff}).
}
 \label{fig:r2n}
\end{figure}

    The equilibrium density minimizes the total energy with respect to the
density, at fixed particle number, $N=\int d{\mathbf r}~n(r)$.  Taking the constraint
of fixed total number of particles into account via a chemical potential, $\mu
= \mu_{\uparrow} = \mu_{\downarrow}$, we find that the first functional
derivative of the energy is
\begin{eqnarray}
    \delta E - \mu\delta N = \int d{\mathbf r}
    \left((3\pi^2)^{2/3}\frac{\hbar^2}{2m}n^{2/3}
    + \frac{1}{2}m\omega^2 r^2 \right. \nonumber \\
      \hspace{50pt}+ \left. \frac{g}{2} n -\mu \right)\delta n = 0,
\label{de}
\end{eqnarray}
which yields the density-position relation,
\begin{equation}
  (3\pi^2)^{2/3}\frac{\hbar^2}{2m}n^{2/3} + \frac{1}{2}m\omega^2 r^2 +
    \frac{g}{2} n =\mu.
\end{equation}
It is most convenient to regard this equation as determining $r^2$ in
terms of $n$:
\begin{equation}
    r^2=-\left((3\pi^2n)^{2/3} -  4\pi|a| n\right)d^4 + R^2(a,N),
 \label{rn}
\end{equation}
where $d=\sqrt{\hbar/m\omega}$ is the oscillator length, and the radius of
the cloud is $R= \sqrt{2\mu/m\omega^2}$.  Equation~(\ref{rn}) can be inverted
either graphically, or by simply solving the cubic equation explicitly.  The
results are shown in Fig.~(\ref{fig:r2n}).  Note that when $a<0$, $r(n)^2$ has
a minimum as a function of density at
\beq
  r_0^2 = -\frac{\pi^2}{12}\frac{d^4}{|a|^2} + R^2,
  \label{r0}
\eeq
where
\beq
  n_0 =  \frac{\pi}{24|a|^3}.
   \label{n0}
\eeq

    In order for the density profile to be stable it must minimize, not
maximize, the energy functional, i.e., the second variation of $E-\mu N$:
\begin{eqnarray}
  \delta^2 E - \mu \delta^2 N  =  \int d{\mathbf r}
   \left((3\pi^2)^{2/3}\frac{\hbar^2}{3m} n^{-1/3} \right.
   \nonumber \\ +\left. \frac{g}{2}\right) (\delta n)^2 ,
  \label{d2e}
\end{eqnarray}
must be positive for all $\delta n$. This condition requires,
\begin{equation}
   k_F(r)|a| < \frac{\pi}{2} \approx 1.57,
    \label{kf}
\end{equation}
at all positions, or equivalently $n \leq \pi/24|a|^3$ (cf.
Eq.~(\ref{n0})).  As we see, Eq.~(\ref{kf}) is the condition that the density
profile corresponds to a minimum of the energy functional (\ref{e_func}).
Note that we have derived this result in terms of the Thomas-Fermi profile
including interactions.  This argument is similar to that of \cite{pethick2}.
The result (\ref{kf}) has also been obtained heuristically by comparing the
repulsive force on an atom arising from the kinetic energy term with the
attractive force due to the mean field interaction energy \cite{gehm}, and
derived via a field-theoretical approach \cite{stoof}.

    When the minimum of $r^2$ vs.  $n$ is at negative $r^2$, the system is
stable because then $n$ is less than $n_0$, so that $k_F|a|$ is always
$<\pi/2$.  On the other hand, when the minimum is at positive values, the
curve does not intersect the horizontal axis, and therefore there is no
solution for the density at the center of the trap.  (An unphysical solution
for the density would have a hole of radius $r_0$ about the center of the
trap.)  However, in this case the density equals $n_0$ at $r=r_0$, and
vanishes as $r\to R$.  Then, $k_F(0)|a|= \pi/2$, and the solution is
marginally stable, not actually minimizing the energy functional (\ref{e_func}).
We conclude that the system is stable when $r_0^2 \leq 0$, and the critical
point is $r_0=0$, corresponding to
\begin{equation}
  R^2= \frac{\pi^2}{12}\frac{d^4}{|a|^2}
\label{R2}.
\end{equation}
and then the density of the system at the center of the trap is given by
Eq.~(\ref{n0}).

    When Eq.~(\ref{R2}) is satisfied the system reaches its maximum number of
particles in equilibrium, given by the area under the $n$ vs $r$ curve,
$N_{max}= 4\pi \int_0^{\infty} n(r) r^2 dr$.  Integrating with respect to the
$n$, we find
\begin{equation}
  N_{max} = \frac{\pi^5}{48}\frac{d^6}{|a|^6}I,
  \label{Nmax1}
\end{equation}
where
\begin{eqnarray}
  I &=& \int_0^1 \left(\frac{2}{3}x^3 -x^2  + \frac{1}{3}\right)^{1/2} (1-x)
    x^4 dx
  \nonumber \\
  & & = \frac{54}{5005} - \frac{40}{9009\sqrt{3}} \approx 0.00823.
\nonumber
\end{eqnarray}
As long as the local density approximation is valid, the condition
(\ref{Nmax1}) is equivalent to the condition (\ref{kf}).  However, the use of
mean field is no longer valid for $k_F|a|\ga 1$.  From (\ref{Nmax1}), the
condition for stable equilibrium is $N^{1/6} \leq 0.612 d/|a|$, which agrees
with that found numerically in Ref. \cite{menotti}.

\begin{figure}
\includegraphics[width=0.47\textwidth]{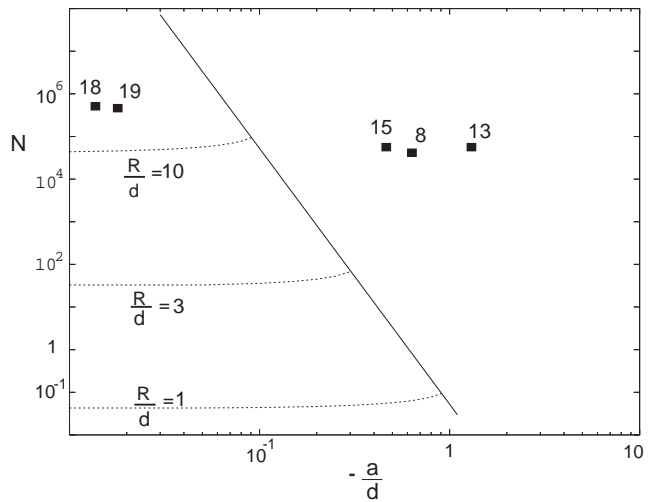}
\caption{The maximum particle number vs.  $-a/d$, as predicted by
Eq.~(\ref{Nmax1}). The nearly horizontal lines are contours of constant
chemical potential in the stable region; here $R/d= \sqrt{2\mu/\hbar\omega}$.
Also shown are the particle numbers  measured in the five experiments in Table
1.  Data points are labelled by the corresponding reference numbers.
}
\label{fig:Nvs_a}
\end{figure}

    Table \ref{exp} lists the number of particles in representative
experiments on $^6$Li gases with two-spin components.  Since these experiments
were not testing the maximum number of particles that could be trapped, they
provide only lower bounds on the maximum number of particles.  For small
$k_F|a| <1$, the experimental particle numbers are well within the bound
(\ref{Nmax1}).  However fermionic systems at larger $k_F|a|$, exceeding
$\pi/2$, do remain stable.  See Fig.  \ref{fig:Nvs_a}.  The argument leading
to Eq.~(\ref{Nmax1}) breaks down at large $k_F|a|$, for then correlations
beyond mean field become important.  In addition, when $n(0)\to n_0$, the
local density approximation fails.  A necessary condition for the local
density approximation to be valid is that the density varies slowly over an
interparticle spacing, or $|\partial \ln n/\partial r| \ll k_F$.  From
Eq.~(\ref{rn}), we thus find the requirement, for $n(0)\simeq n_0$, that
$1-2k_F|a|/\pi \gg (|a|/d)^2$ for the local density approximation to be
valid, where we have taken $r^2$ to be bounded by the critical $R^2$ in
Eq.~(\ref{R2}).  This condition clearly breaks down as $n(0)\to n_0$, where
$dn/dr$ diverges, as one sees in Fig.~\ref{fig:r2n}.

\begin{table}[htbp]
\caption{Experimental measurements of particle numbers in $^6$Li gases.  In
the regime
$k_F|a|<1$, Eq.(\ref{Nmax1}) is satisfied.}
$$
\begin{array}{clccrcc}
\hline\hline
&\text{Expt.} & T /T_F & -a/a_0 & N_{obs} & N_{max}&  k_F|a| \\
\hline
& \text{\cite{jochim}}    & <0.2      & 3500  & 9\times10^5      & \sim 10^{12} &  < 1        \\
& \text{\cite{kevin}}     & <0.1      & \sim500   & 6\times10^5      & \sim 10^9    & << 1        \\
& \text{\cite{partridge}} & \sim 0.75 & 15600 & 1.3-9\times 10^4 & \sim 7       & >>1         \\
& \text{\cite{gehm}}     & \sim 0.15 & 10^4  & 8.0\times 10^4   & \sim 1       & \approx 2.5 \\
& \text{\cite{ohara}}     & \sim 0.1  & 10^4  & 7.5\times10^4    & \sim 0       & \approx 7.4 \\
\hline\hline
\end{array}
$$
\label{exp}
\end{table}

\section{Near the unitary regime}

    When $k_F|a|$ becomes large perturbation methods fail, and the question of
stability is more involved.  In the unitarity regime ($k_F|a|\to\infty$), the
energy of the ground state is bounded and can be written in the scale-free
form \cite{henning,carlson,baker,bourdel} $E_0 = (1 + \beta) E_{FG}$, where
$\beta$ is universal constant $\sim -0.5$.  To approach the problem, we
interpolate the energy between small $|a|$ and $-\infty$ by calculating in
mean field theory with a density dependent effective scattering length,
\begin{equation}
  a_{\rm eff}= \frac{a}{1-\gamma k_F a},
   \label{aeff}
\end{equation}
for $k_F a<0$, and replacing the coupling $g$ in Eq.~(\ref{e_func}) by the
local coupling $g_{\rm eff}=4\pi\hbar^2 a_{\rm eff}/m$.  The choice $\gamma =
-20/9\pi\beta$ reproduces the known energies in the weakly interacting limit
and in the unitary regime \cite{delta}.

    Repeating the above Thomas-Fermi calculation with this density-dependent
scattering length, we find, first, the position as a function of density,
\begin{eqnarray}
    r^2=-\left[k_F^2 -  4\pi|a_{\rm eff}| n\left(1-\frac{\gamma}{6}
   k_F|a_{\rm eff}|\right) \right]d^4 + R^2.
 \label{rn1}
\end{eqnarray}
The density profile is similar to that of the stable configurations shown
in Fig.~2.  We show in that figure, as a dashed line, the density profile
calculated with Eq.~(\ref{rn1}) $a = -0.5$, and $\beta = -0.5$.

\begin{figure}
\includegraphics[width=0.47\textwidth]{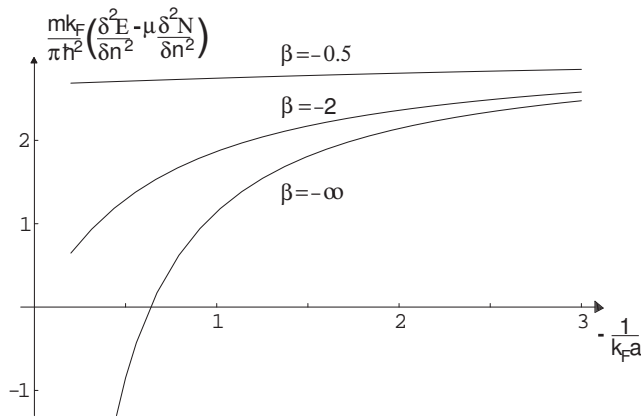}
\caption{The second variation of $E-\mu N$, Eq.~(\ref{d2E_eff}), as a
function of $-1/k_F|a|$, calculated with the effective scattering length,
$a_{\rm eff}$, Eq.~(\ref{aeff}).}
 \label{fig:d2E_vs_ka}
\end{figure}

    The second functional derivative of the energy with respect to $n$ is,
\begin{eqnarray}
    \frac{\delta^2 E - \mu \delta^2 N}{(\delta n)^2} = \frac{\pi \hbar^2}{m
    k_F}\big(\pi + 2 k_F a_{\rm eff} +\frac{10}{9}\gamma (k_Fa_{\rm
    eff})^2
    \nonumber\\
     + \frac{2}{9}\gamma^2(k_Fa_{\rm eff})^3
   \big).
\label{d2E_eff}
\end{eqnarray}
In the limit $\gamma k_F|a|<<1$, $a_{\rm eff}\sim a$, and
Eq.~(\ref{d2E_eff}) predicts that the gas is stable for all $k_F|a|$, in
distinction to the prediction of Eq.~(\ref{kf}).  In the opposite limit,
$\gamma k_F|a|>>1$, $|a_{\rm eff}|\to 9\pi\beta/20k_F$, a constant, and
$\left(\delta^2 E - \mu\delta^2 N\right)/(\delta n^2)=(\pi^2\hbar^2)/(m
k_F)(1+ \beta/2)$.  Hence the gas is stable for $\beta>-2$, as can be seen in
Fig.~\ref{fig:d2E_vs_ka}, a plot of $(m k_F/\pi\hbar^2 )(\delta^2 E -
\mu\delta^2 N)/(\delta n)^2$ as a function of $-1/k_F|a|$ for various $\beta$.
Experiments on $^6$Li find\cite{gehm,bourdel,partridge2}, $\beta\sim-0.5$;
for these values, the second variation is always positive, predicting that
the gas should be stable for all $k_F|a|$, in agreement with experiment.

\section{Even higher density}

    In the scale-free unitary regime the interparticle spacing is large
compared with the range of the potential, and the system is stable.  However,
at higher densities when the interparticle spacing becomes comparable to the
range of the potential, the system can in fact become unstable, as we see from
the following model calculation.  Let us assume that the interparticle
interaction is an attractive square well potential with range $r_0$ and depth
$V (<0)$.  The energy calculated assuming a Hartree-Fock wave function is a
rigorous upper bound to the exact total energy, $E$; thus
\begin{equation}
    \frac{E}{N} \le \frac{3\hbar^2 k_F^2}{10m}+ \frac{2\pi r_0^3}{3}n V.
\end{equation}
As $n$ increases, the energy becomes unbounded below; the kinetic energy
does not save the system from collapse.  We note that at $V= -\hbar^2\pi^2/4m
r_0^2$ the two body-scattering amplitude has a resonance, the scattering
length diverges, and the interaction energy per particle becomes $E_{\rm
int}/N= -(5\pi/54)k_F r_0 E_{FG}$.  The instability in this model should not
occur, however, for realistic interatomic potentials which have a repulsive
core.

    We thank Randy Hulet and Zhenhua Yu for useful discussions.  This work
was supported in part by NSF Grants PHY03-55014 and PHY05-00914.


\begin{thebibliography}{99}
\bibitem{delta} Within two body physics, the interaction energy is more
correctly given in terms of two-body phase shifts, rather than scattering
lengths:  $E_{\rm int} \simeq -4\pi\hbar^2\delta n/km$, where $k \la k_F $ is
an averaged relative momentum in the interaction, $\sim k_F$, and $\delta =
-\tan^{-1}(ka)$ is the scattering phase shift.  Thus in this case, we would
write $ka_{\rm eff} \sim \tan^{-1}(ka)$, which is similar in structure to
Eq.~(\ref{aeff}).  This result corresponds to $\gamma = 2k/\pi k_F$.
\end{thebibliography}
\end{document}